# The Impact of Heterogeneous Shared Leadership in Scientific Teams

Huimin Xu
*School of Information, University of Texas at Austin, Austin, TX 78701, USA*
Meijun Liu
*Institute for Global Public Policy, Fudan University, Shanghai 200433, China*
Yi Bu
*Department of Information Management, Peking University, Beijing 100871, China*
Shujing Sun
*School of Management, University of Texas at Dallas, Dallas, TX 75080, USA*
Yi Zhang
*Faculty of Engineering and IT, University of Technology Sydney, NSW 2007, Australia*
Chenwei Zhang
*Faculty of Education, University of Hong Kong, Hong Kong 999077, China*
Daniel E. Acuna
*Department of Computer Science, University of Colorado at Boulder, Syracuse, Boulder, USA*
Steven Gray
*School of Business, University of Texas at Austin, Austin, TX 78712, USA*
Eric Meyer
*School of Information, University of Texas at Austin, Austin, TX 78701, USA*
Ying Ding
*School of Information, University of Texas at Austin, Austin, TX 78701, USA*

**Corresponding author: Ying Ding (ying.ding@ischool.utexas.edu).**

# The Impact of Heterogeneous Shared Leadership in Scientific Teams

**Abstract:** Leadership is evolving dynamically from an individual endeavor to shared efforts. This paper aims to advance our understanding of shared leadership in scientific teams. We define three kinds of leaders, junior (10-15), mid (15-20), and senior (20+) based on career age. By considering the combinations of any two leaders, we distinguish shared leadership as "heterogeneous" when leaders are in different age cohorts and "homogeneous" when leaders are in the same age cohort. Drawing on 1,845,351 CS, 254,039 Sociology, and 193,338 Business teams with two leaders in the OpenAlex dataset, we identify that heterogeneous shared leadership brings higher citation impact for teams than homogeneous shared leadership. Specifically, when junior leaders are paired with senior leaders, it significantly increases team citation ranking by 1-2%, in comparison with two leaders of similar age. We explore the patterns between homogeneous leaders and heterogeneous leaders from team scale, expertise composition, and knowledge recency perspectives. Compared with homogeneous leaders, heterogeneous leaders are more adaptive in large teams, have more diverse expertise, and trace both the newest and oldest references.

Introduction

The complex challenges of scientific discovery, render us to rethink whether the traditional sole leadership is still suitable for scientific teams, or if shared leadership where multiple leaders engage in leadership activities is more desirable to creative tasks. Internally, the demands of the scientific workforce are different. Researchers are a highly educated workforce who contribute intellectual capital. These knowledge workers are not merely satisfied with material wealth (e.g., money), but also have emotional and professional pursuits, such as leadership impact (Pearce et al., 2004). The absolute authority of a sole leader risks stifling the creativity of scientists - especially those coming from different domains with drastically varied norms, practices, and thought processes. Drawing on social identity theory, Hogg (2008) emphasized the identity dimension of leadership. In innovative work contexts, group belongingness can foster individuals carrying different knowledge to transform personalized identity to serve the collective creative identity as a means to maintain positive self-images in groups (Hirst et al., 2009). The innovation of leaders would directly motivate the creative efforts of followers in the whole team (Hirst et al., 2009). Externally, scientific teams need to keep pace with the rising trend of interdisciplinarity and internationalization. Interdisciplinary projects involve experts from different fields and require greater division of labor (Haeussler & Sauermann, 2020; Zhang et al., 2018; Zhou et al., 2022). International cooperation brings higher transaction costs, language barriers, and cultural differences, which would hinder research creativity (Wagner et al., 2019). Multiple leaders, however, can exhibit their respective areas of skills, expertise, and abilities to evolving tasks (Pearce & Conger, 2002). In scientific publications, the trend that multiple leaders collectively manage the scientific teams and share the leading power is rising. A signal is that the percentage of equal first authors and corresponding authors increases linearly in scientific publications (Hu, 2009). But only a handful of current literature contributes to the understanding of shared leadership in scientific team innovation and impact.

Division of labor in knowledge production determines leading and supporting roles. By analyzing self-reported contributions to publications, Xu et al. (2022a) measured the number of leaders with keywords (e.g., "design," "lead," "supervise") in proportion to team size. It was found that flat team structures with more leaders were conducive to disruptive innovation and long-term impact. Notably, however, this research studied the role of shared leadership, yet treated all leaders the same, ignoring similarities and differences between individuals. Instead, it is important to frame seniority levels around distinctions in different leadership skills (Mumford et al., 2007). The order of authorship is another common manifestation of leadership (Chinchilla-Rodríguez et al., 2019; Liu et al., 2014). Traditional literature usually takes the first and last authors as leading authors, which is a form of shared leadership. First authors and last authors typically contribute more than middle authors in publications in all disciplines (Larivière et al., 2016). In experimental research, first authors conventionally are the students being supervised by last authors (Costas & Bordons, 2011). Young scholars who perform most of the technical parts tend to be listed as first authors, whereas senior scholars who conceive the conceptual design are usually listed last (Larivière et al., 2016). In this sense, the first authors do not fit the definition of scientific leadership that leaders have the ability to obtain resources, have the expertise to initiate and develop projects, and have higher impact and productivity (Chinchilla-Rodríguez et al., 2019). Career age is a direct indicator of contribution. Senior researchers tend to shoulder leadership responsibility. Drawing on Merton's principle of cumulative advantages (1973), the gap between senior researchers and junior ones is exacerbated since senior authors are more productive and are more cited in publications, have accumulated enough experience and knowledge in research, and have the ability to provide rich funds and advanced equipment to support research. As authors shift from junior to full careers, they shoulder more responsibilities in leading teams than supporting them (Larivière et al., 2016). However, current research seldom distinguishes scientific leaders with different career ages and their roles in shared leadership.

Shared leadership also has shortcomings, namely low efficiency (Pearce, 2004), diffusion of responsibility (Zhu et al., 2018), and competition and conflicts (Weiss et al., 1992). O'Toole et al. summarized that the success of the co-leader model lies in "how complementary the skills and emotional orientations and roles of the leaders are" (2002, p. 71). Distinguishing heterogeneous leaders and homogeneous leaders is key to understanding the advantages and disadvantages of shared leadership in science. Diversity and complementarity from heterogeneous leaders strengthen shared leadership, including handling complex tasks (Carpenter & Fredrickson, 2001). As noted above, the role overlap from homogeneous leaders can weaken shared leadership. Demographic-based characteristics (e.g., age) are crucial aspects of heterogeneity and homogeneity in team leaders (Sperber & Linder, 2018). In scientific teams, when leaders in the same career age cohorts are paired, they are homogeneous leaders. Otherwise, they are defined as heterogeneous leaders. How heterogeneous leaders and homogeneous leaders influence team scale, expertise diversity, knowledge recency, and ultimate performance is seldom explored in scientific teams.

This paper aims to study the relationships between shared leadership and team performance in creative scientific discoveries, specifically highlighting heterogeneous and homogeneous shared leadership to identify unique patterns. This study addresses aforementioned research gaps by taking Computer Science (CS), Sociology, and Business as test fields to compare heterogeneous and homogeneous shared leadership in team performance. This paper is structured as the followings. Section 2 highlights the related literature. Section 3 outlines the research methods. Section 4 details the findings and provides comparisons and discussions. Section 5 concludes the study and points out future research directions.

## Literature Review

### *Heterogeneous vs Homogeneous Shared Leadership*

Researchers are committed to studying the heterogeneity/homogeneity of leaders in creative teams as manifested through variables such as age, expertise, character (O'Toole, 2002), tenure, educational and functional composition (Hambrick et al., 1996), leadership style (Zhu et al., 2011), and status (Watts, 2010). In US airline companies, where leaders are heterogeneous in function, tenure, and educational background, they have great vision and external connections to initiate creative frequent-flyer programs and expand new markets (Hambrick et al., 1996). But heterogeneity can produce conflicts between leaders and thus impede the response process. In the online knowledge production community, heterogeneous leaders with different leadership styles could shoulder different division (Zhu et al., 2011), where transformational leaders are socially oriented and person-focused, and transactional leadership is task-driven, offering rewards for followers who are compliant but punishment for those who are deviant (Bass et al., 2003). In scientific discovery, Watts (2010) indicated that heterogenous leadership (e.g., a novice leader paired with an experienced one) has more potential benefits than homogenous leadership (e.g., leaders with equal status). Firstly, this is beneficial to create an environment where leaders can respect and learn from each other "characterized by intellectual generosity" (Watts, 2010, p. 336). Secondly, the heterogeneity of shared leadership can encourage student-centered supervisory ways (Firth & Martens, 2008; Phillips & Pugh, 2015). Thirdly, leaders' heterogeneity can advance interdisciplinarity (Haythornthwaite, 2006). Harmonious relationships, however, will be broken in some cases. For example, when two leaders have competing ideas or conflicts, a student might feel confused, distracted, or contradictory. Students are more likely to see their leaders separately rather than together. This will increase the difficulty of reaching a consensus. However, the discussion about the advantages and disadvantages of homogeneous/heterogeneous leaders is only limited to small-scale groups in qualitative methods. The question of how to scale up the results within millions of scientific teams is still a new direction.

*Shared Leadership and Team Scale*

Literature in team science always highlights differences between large and small teams. Large teams enable a clear division of labor and specialization in solving complex problems, particularly when there is a high degree of certainty and independency about the tasks (Meyer & Schroeder, 2015; Whitley, 2000). Specifically, by studying the collaboration practices of sociologists, Hunter & Leahey (2008) concluded quantitative methods were more suitable for large teams since they enabled easily-divisible tasks, like collecting, cleaning, coding, and analyzing data. Meanwhile, large teams have extra cognitive resources because they can be more diverse in skills and knowledge than small teams (Bantel & Jackson, 1989). Larivière et al. (2015) analyzed team size using three indicators: the number of authors, institutions, and countries. They found that as teams become larger and more diverse, they can receive more influence in citations. However, large teams are more likely to produce conflicts due to increased communication costs and decreased support (Staats et al., 2012). Although large teams can generate more ideas than small teams in experimental brainstorming settings, being exposed to too many ideas can lead to distraction and information overload (Paulus et al., 2013). By analyzing millions of teams in paper publication, patent invention, and software development, Wu et al. (2019) suggested that small teams can achieve disruptive innovation, although they do not receive much attention, compared with the incremental innovation in large teams. Current research focuses more on the comparison between small and large teams, but how to explore the roles of shared leadership in these teams with distinct sizes is still unknown.

*Shared Leadership and Expertise Diversity*

Researchers always take team members' expertise into consideration while discussing collaboration. In scientific teams, the most common methods to mine research expertise are based on keyword extraction or topic modeling in papers. The main conclusion of this line of research is that researchers' expertise homophily is much easier to boost collaboration but does not necessarily lead to better performance. Kraut et al. (2014) calculated the similarity between authors by embedding their publications' abstracts into a semantic space, and concluded that authors sharing similar expertise are more likely to collaborate. Likewise, Ding (2011) found that productive authors in the information retrieval field tend to share similar research expertise with collaborators identified by the topic modeling method. Extending the top 100 productive authors to all authors in the information retrieval fields, Zhang et al. (2016) suggested common research interests are not the dominating factor in collaboration, and authors with different research interests can work together. Bu et al. (2018b) explored the relationship between authors' impact and their collaborators' topic diversity in computer science. They concluded authors who have more impact (i.e., high h-index) tend to collaborate with authors with diverse research interests. However, high expertise overlapping does not boost communication and division of labor, especially when all members share the same language and a joint project goal (Giuri et al., 2010). Even though the science of science researchers consider the topic diversity within collaborators, they do not discuss the expertise composition of multiple leaders.

*Shared Leadership and Knowledge Recency*

A paper's references represent a team's accumulated knowledge about a specific topic. Team members of different career ages contribute differently to the knowledge flow (Zou et al., 2023), such as the citing practice. There is an age-stratified difference in accepting new ideas (Merton, 1973). Cui et al. (2022) generalized this finding in all fields in the MAG dataset. By uncovering individual scientists' careers, they found that aging scientists favored citing older references, and their ideas stayed in their original state when they were young. What is worse, by classifying citations into constructive or contrasting, they found senior researchers even challenged new ideas. Aging brings social and cultural resistance. Kuhn proposed that science advances in a revolutionary rather than an incremental way (1970). These scientific revolutions are started by junior scientists who are very young and new to the field. But the glory and shine of star scientists might prevent newcomers and outsiders from challenging their authority (Azoulay et al., 2019). Azoulay et al. (2019) found that the premature death of superstar life scientists can lead to a marked increase in outsiders who contribute to the field. The older people resist the younger because they are restricted in "response to innovation by … substantive and methodological preconceptions and by … other cultural accumulations" (Barber, 1961, p. 601). A higher proportion of older researchers tend to lead to older citations (Barnett & Fink, 2008). But working with young scholars could help them slow the aging rate and cite more recent references, thus boosting scientific advancement (Cui et al., 2022). Whether these conclusions similarly applied to shared leadership still needs to be tested.

## Methods

*Dataset*

OpenAlex is a scientific knowledge graph, including papers, disambiguated authors, venues, institutions, and discipline information (Priem et al., 2022). Disciplines have a hierarchical structure with 6 levels, where the roots have 19 disciplines. We chose the root-level discipline of Computer Science and got 33,674,530 CS papers in journals and conference proceedings from 1800 to 2023. Given that we want to analyze shared leadership in teams, we exclude papers with sole author and keep papers with at least two leaders whose career ages are above 10 (see Measures for leader definition). There are 4,899,720 papers with multiple leaders left. Among these shared leadership papers, papers with two leaders occupy 60.7% (2,973,463). Thus, we take the most typical two-leader combinations to represent shared leadership in this paper. To choose the leaders, we not only control career ages above 10, but also restrict the positions in the authorship. By tracing the publications of ACM (The Association for Computing Machinery) fellows, Fernandes et al. (2022) concluded that these CS leaders typically place their names on the right side of the bylines, say, last positions. Besides the first and last positions, the second-to-last position is the most important position of the paper, which suggests seniority and leadership (Helgesson & Eriksson, 2019). Hence, we select shared senior authors (above 10) in the first and last positions or the last two, given that leaders usually occupy these positions. To generalize our results in CS, we choose another two social science disciplines, Sociology and Business. Finally, there are 1,845,351 CS, 254,039 Sociology, and 193,338 Business teams with two leaders left.

*Measures*

OpenAlex deals with the author name disambiguation in papers and provides the unique id of each author (Priem et al., 2022), thus we can trace the career age of each author after he/she published the first paper. For example, an author published his/her first paper in 2000, then his/her career age was 5 in 2005. Milojević et al. (2018) defined different researchers based on career age, from junior (1–10), early-career (11–15), mid-career (16–20) to full-career scientists (>20). Bu et al. (2018a) defined milestones of CS researchers that they usually go through Ph.D. studies in the first five years, postdocs or assistant professors in the following five academic years, then assistant/associate professors (Year 10-15), next associate/ full professor (Year 15-20), finally finish their career. Based on the definition of different leaders (Bu et al., 2018a; Larivière et al., 2016; Milojević et al., 2018), we firstly define leaders as those researchers whose career age are above 10, which suggests that they have accumulated 10-year research experience. Then, we distinguish three kinds of leaders, junior leader (>=10 & <15), mid leader (>=15 & <20), and senior leader (>=20). To differentiate shared leadership with two leaders, there are six different combinations among these three kinds of leaders, two junior leaders (2,0,0), two mid leaders (0,2,0), two senior leaders (0,0,2), one junior leader, and one mid leader (1,1,0), one mid leader and one senior leader (0,1,1), and one junior leader and one senior leader (1,0,1). Leaders within the same age cohort are defined as homogeneous leaders, whereas leaders from different age cohorts are heterogeneous leaders. Meanwhile, we limit these two leaders' positions to the first-last or the last two positions in publications. We describe the number and median value of career age difference between two leaders in each combination.

Table 1. The descriptive variables of leader categories in Computer Science

| Leader Categories | Number | Median Value for Difference of Leaders' Career Age |
|---|---|---|
| (2, 0, 0) | 243,204 | 1.0 |
| (0, 2, 0) | 126,808 | 1.0 |
| (0, 0, 2) | 314,535 | 4.0 |
| (1, 1, 0) | 314,275 | 5.0 |
| (0, 1, 1) | 353,554 | 8.0 |
| (1, 0, 1) | 492,975 | 12.0 |

To control confounding variables in the relationship between leaders and team impact, we use the multivariable regression below (Equation 1 & 2):

$$Team\ Impact_i = \alpha + \beta_1(Leader\ categories_i) + \beta_2(Controls) + e_i \quad (1)$$
$$Team\ Impact_i = \alpha + \beta_3(Difference\ of\ Leaders'\ career\ age_i) + \beta_2(Controls) + e_i \quad (2)$$

The dependent variable $Team\ Impact_i$ is the value for the 2-year or 5-year citation percentile of team i. The independent variables are $Leader\ Categories_i$ in the equation 1 and $Difference\ of\ Leaders'\ career\ age_i$ in the equation 2. There are six different leader categories shown in Table 1. $Difference\ of\ Leaders'\ career\ age_i$ means the career age difference of two leaders within a team. For example, the senior leader's career age is 20, and the junior leader's career age is 11, then the difference is 9. To control the time-invariant factor, the publication year is treated as a fixed effect. $\beta_1$ and $\beta_3$ are the coefficients of independent variables. $\beta_2$ represents the coefficients for control variables, e.g., team size. $e_i$ represents the residual, the difference between actual values and predicted values. $\alpha$ is the constant term, the mean value of the dependent variable, when all the independent variables in the model are zero.

*Evaluation*

To evaluate the validity of our shared leadership measurement, we match the leaders we identified in publications with their self-report contributions. Firstly, we downloaded self-reported contribution data[1] for PLOS ONE, PNAS, Science, and Nature. Then, through doi, the unique paper identification, we matched 2,379 papers (1,626 CS, 710 Sociology, and 43 Business) in the OpenAlex shared leadership dataset and in the contribution dataset. By identifying the keywords in contributions, "conceive", "design", "lead", "supervise", "coordinate", "interpret", and "write", we selected researchers who shouldered the leadership roles (Xu et al., 2022a). In these matched papers, these keywords separately appear 4,502 ("conceive"), 6,290 ("design"), 36 ("lead"), 55 ("supervise"), 10 ("coordinate"), 67 ("interpret"), 6,576 ("write") times. Please note that one researcher can shoulder multiple leading roles. Finally, leaders whose career ages are above 10 and whose authorships are in the first or last two positions, also reported their leadership contributions in 1,871 papers (1,273 CS, 561 Sociology, and 37 Business). The accuracy rate of our measurement is 78.6%. Table 2 describes the accuracy rate in different leader categories.

Table 2. The accuracy rate in different leader categories

| Leader Categories | Total Number | Correct Number | Accuracy Rate |
| --- | --- | --- | --- |

---

[1] https://zenodo.org/record/6569339#.Y8n3TuzMK0c

| | | | |
|---|---|---|---|
| (2, 0, 0) | 292 | 226 | 0.77 |
| (0, 2, 0) | 137 | 103 | 0.75 |
| (0, 0, 2) | 452 | 364 | 0.81 |
| (1, 1, 0) | 385 | 293 | 0.76 |
| (0, 1, 1) | 679 | 552 | 0.81 |
| (1, 0, 1) | 434 | 333 | 0.77 |

## Results and Discussion

*Heterogeneous vs Homogeneous Shared Leadership*

Fig 1a is an illustrated case of an independent team where two leaders and two students collaborate. Fig 1b shows six different leader combinations where two leaders in shared leadership belong to different junior, mid, and senior categories. Leaders belonging to the same age cohort are homogeneous teams, (2,0,0), (0,2,0) and (0,0,2), whereas leaders belonging to the different age cohorts are heterogeneous teams, (1,1,0), (1,0,1) and (1,0,1). We want to observe the citation patterns in heterogeneous and homogeneous shared leadership. To reduce the citation inflation effect, we calculate the 5-year citation percentile for all papers published in the publication year. It is a relatively fair way to compare citations for papers published in different years. The high (low) citation percentile means high (low) citations in the year, scaling from 0 to 100. For instance, the 5-year citation percentile of a paper published in 2013 is 99%, which means this paper belongs to the top 1% of most frequently cited publications in 2013. Meanwhile, the distribution of leader categories is not even (Table 1), and the number of homogeneous leader combinations, such as (0,2,0), (2,0,0), are less than heterogeneous leader combinations. To reduce the effect of uneven frequency distribution, we randomly chose 50,000 teams from each category ten times and then calculated the average citation percentile. The results in Fig 1c show that heterogeneous shared leadership has a higher citation percentile than homogeneous shared leadership within 5 years on average. Besides observing the difference in categories, we can also extend the categorical variables to continuous variables. Career age serves as a proxy of power since it represents experience, resources, and wisdom (Xu et al., 2022b). The career age difference between the two most senior leaders corresponds to the measure of power distance between the most powerful person and the next powerful person (Eisenhardt & Bourgeois, 1988). When we calculate the specific career age difference between two leaders, we find that the larger the age difference in leader combinations is, the higher the citation percentile is (Fig 1d). It also suggests that the heterogeneity of leaders brings more citations. We choose 1,117,662 papers where the first author is a junior researcher (career age <10) among those with two leaders. To clearly observe the difference between homogeneous leaders and heterogeneous leaders, we normalize the mean 5-year citation percentile with z-score methods, $z = (x-\mu)/\sigma$, where x is the raw value, μ is the mean value of the population, and σ is the standard deviation value of the population. We conclude that heterogeneous leaders are better than homogeneous leaders in paper citations when we control the career age of students from 1 to 10 (Fig 1e). This finding is consistent with the conclusions from Watts (2010) that the assembly of junior and senior leaders outperforms leaders of similar age in thesis supervision. Heterogeneous leaders in scientific teams can bring complementarity and diversity to handle ambiguous, creative, complex, and interdependent tasks (Carpenter & Fredrickson, 2001; Hambrick et al., 1996).

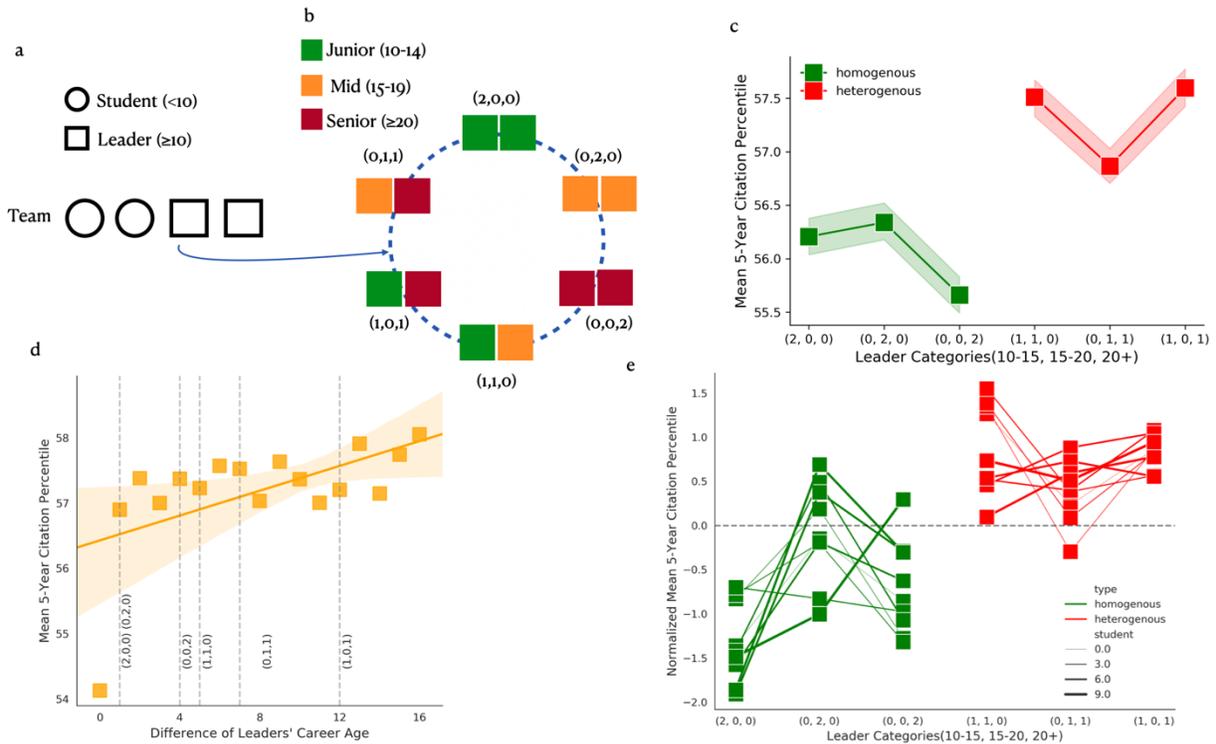

Fig 1. **a**. An example of a team consisting of two students (career age less than 10) in circle shape and two leaders (career age greater than or equal to 10) in square shape. **b.** Distinguish leaders with junior (10-15, green), mid (15-20, orange) and senior (20+, red), and categorize two

leaders into six leader combinations. **c.** Heterogeneous leaders (red) in career age have more citations than homogeneous leaders (green) in CS. We calculate papers' citation percentile and then aggregate the mean value in the six different leader categories within 5 years. Bootstrapped 95% confidence intervals are shown as translucent bands. **d.** Larger difference in career age within leaders brings higher citations in CS. We calculate papers' citation percentile in the publication year and then aggregate the mean value based on the difference of leaders' career age (from 0 to 16 in terms of integers) within 5 year. We fit these mean values with a linear regression model. Bootstrapped 95% confidence intervals are shown as translucent bands around the regression line. We label the median value for six leader categories of leaders' career age difference with grey dashed line. **e.** Leaders' influence for different young scholars aging form 1 to 9 in CS. We calculate papers' citation percentile in the publication year and then aggregate the mean value in six different leader categories based on papers' first author age within 5 year. The gradient color of lines represents the scale of students' career age, darker lines means higher career age. We normalize the mean 5-year citation percentile with Z-score methods. The grey dashed line clearly differentiates the citation difference of homogeneous leaders (green) and heterogeneous leaders (red).

To generalize the results in CS, we choose another two disciplines, sociology and business. We found that teams with heterogeneous leaders (red) consistently have higher citations than those with homogeneous leaders (green). Still, there are some differences across these disciplines. For example, when leaders are homogeneous in career age, Business teams with two senior leaders (2,0,0) perform best, whereas CS teams with two senior leaders (2,0,0) perform worst. When leaders are heterogeneous, CS teams with one junior leader (1,0,1) and (1,1,0) have a higher citation rank than teams without a junior leader (0,1,1). In contrast, sociology and business teams with one senior leader (1,0,1) and (0,1,1) rank higher than teams without one senior leader (1,1,0).

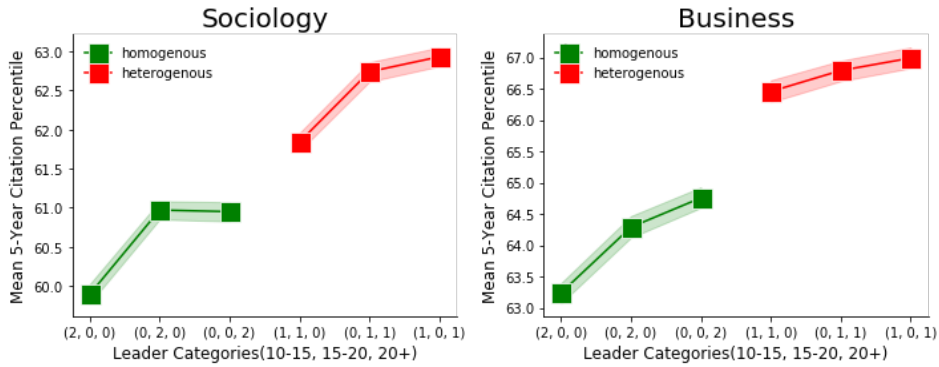

Fig 2. Heterogeneous leaders (red) in career age have more citations than homogeneous leaders (green) in Sociology and Business.

Our definition of junior, mid, and senior leaders is based on career ages 10, 15, and 20. The starting point of becoming a leader is 10 in career age, and the transition period from a lower leader to an upper one is 5. To confirm that our results are robust in different subdivisions. We change the starting point from 10 to 7 (Fig 3a), 13 (Fig 3b), 15 (Fig 3c), and the transition period from 5 to 7 (Fig 3d). Overall, heterogeneous leaders (red) have higher citations than homogeneous leaders (green).

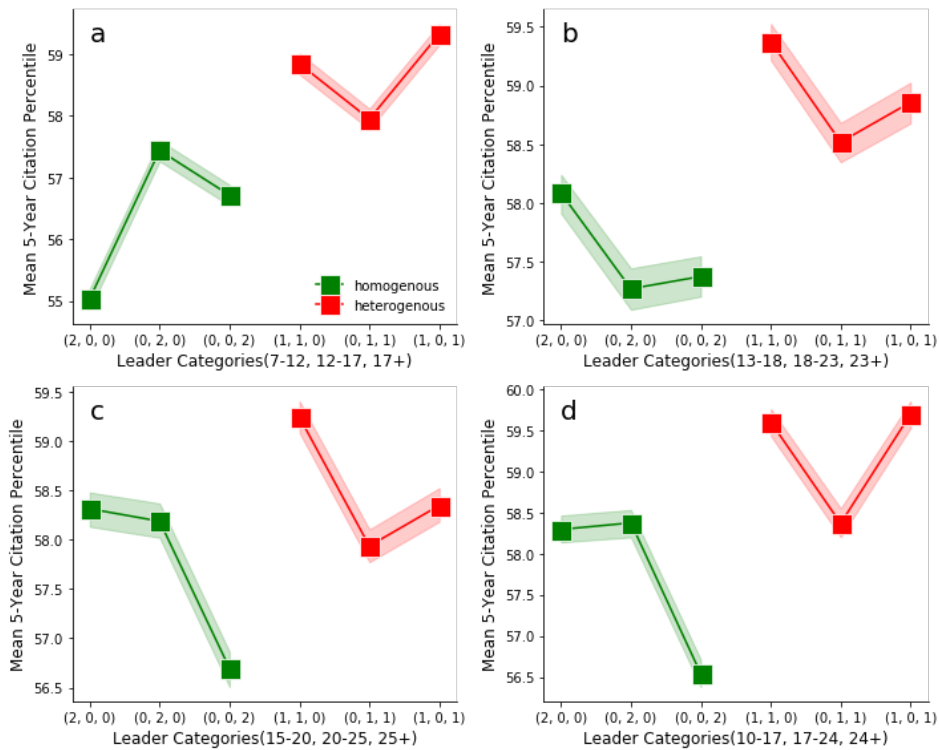

Fig 3. Robust Testing in distinguishing leaders based on different criteria in CS. **a**. junior (7-12), mid (12-17), senior (17+) **b**. junior (13-18), mid (18-23), senior (23+) **c**. junior (15-20), mid (20-25), senior (25+) **d**. junior (10-17), mid (17-24), senior (24+) leaders in six categories

In Table 3, the coefficient values $\beta_1$ in the equation 1 increases in five leader categories within 2-year citation percentile (model 1) and 5-year citation percentile (model 3), compared with the (2,0,0) category, except the (0,0,2) category. It suggests that the (0,0,2) category has the lowest citation percentile, which further confirms the results in Fig 1c. These categories are ranked based on the median value of leaders' career age difference (Table 1). In the models (2) and (4), there is also a significant positive correlation ($\beta_3$ = 0.04 in equation 2) between the difference in leaders' career age and team impact (P < 0.001). When we control team size and the year fix effect, the results are still consistent. Larger teams can bring the higher citation percentile through positive $\beta_2$ in models 1-4.

Table 3. Multivariable regression for leaders and team impact in CS

|  | 2-year citation percentile | | 5-year citation percentile | |
|---|---|---|---|---|
|  | (1) | (2) | (3) | (4) |
| Leader Categories |  |  |  |  |
| (0, 2, 0) | 0.61*** |  | 0.56*** |  |
|  | (0.09) |  | (0.09) |  |
| (0, 0, 2) | 0.02 |  | -0.24** |  |
|  | (0.07) |  | (0.07) |  |
| (1, 1, 0) | 1.75*** |  | 1.88*** |  |
|  | (0.07) |  | (0.07) |  |
| (0, 1, 1) | 1.21*** |  | 1.11*** |  |
|  | (0.07) |  | (0.07) |  |
| (1, 0, 1) | 1.84*** |  | 1.85*** |  |
|  | (0.07) |  | (0.07) |  |
| Difference of Leaders' Career Age |  | 0.04*** |  | 0.04*** |
|  |  | (0.002) |  | (0.002) |
| Team size | 0.29*** | 0.30*** | 1.69*** | 0.14*** |
|  | (0.01) | (0.01) | (0.03) | (0.01) |
| Fixed effect: Year | Yes | Yes | Yes | Yes |
| $R$ | 0.002 | 0.001 | 0.002 | 0.001 |
| $N$ |  | 1,845,351 |  |  |

Note. * $P < 0.05$; ** $P < 0.01$; *** $P < 0.001$; 2-year means 2-year citation, 5-year means 5-year citation

*Shared Leadership and Team Scale*

In Fig 4a, we divide teams into large teams (more than 3 people) and small teams (less than or equal to 3 people). Large teams receive more attention than small teams. In large teams, heterogeneous teams receive more citations than homogeneous teams. It suggests that heterogeneous leaders with both junior and senior leaders are more adaptive in large teams. In small teams, homogeneous teams have a sharp decrease in the mean 5-year citation percentile with increased leaders' career age. Especially for teams with two senior leaders, they are not suitable for leading small teams. This further proves that senior leaders are more suitable to large teams since they have visioning, personnel resources, and financial resources (Mumford et al., 2007) but do not necessarily boost more effective communication and adaptability in small teams (Katzenback & Smith, 1993).

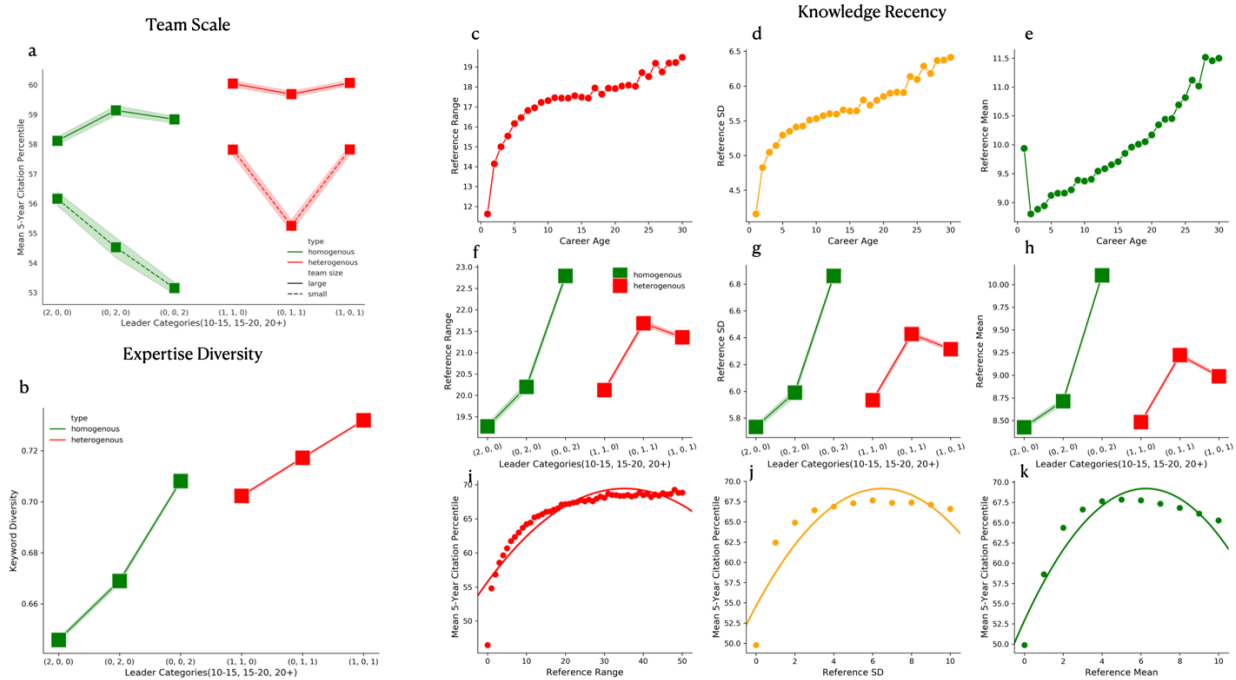

Fig 4. In CS field, **a.** Heterogeneous leaders in large teams have more citations than homogeneous leaders in larger teams. Senior leader combinations in homogeneous leaders have the least citations. We calculate papers' citation percentile and then aggregate (mean value) in the six different leader categories in different teams within 5 year. Bootstrapped 95% confidence intervals are shown as translucent bands. **b.** Homogeneous leaders have lower expertise diversity than heterogeneous leaders. We calculate the ratio between the overlap keyword sets number and all the keyword sets number for two leaders until they publish the focal paper. **c-e.** Researchers who are older in career age have larger reference range (c), reference standard deviation (d) and reference mean (e). Each point in the picture represents the mean values of reference variables grouped by the career age value. **f-h.** Papers with senior leaders have oldest reference, whereas papers with junior leaders have most recent references. We calculate papers' reference range (f), reference standard deviation (g) and reference mean (h) and then aggregate (mean value) in the six different leader categories. Bootstrapped 95% confidence intervals are shown as translucent bands. **i-k.** Curvilinear relationship between reference aging and paper impact. We calculate papers' citation percentile in the publication year and then aggregate (mean value) based on papers' reference range, standard deviation and mean values within 5 years.

*Shared Leadership and Expertise Diversity*

We use the keywords of each paper to represent the authors' expertise. For each leader in each paper, we trace all the keywords of one's publications before publishing the focal one. For two leaders in each paper, we measure the percentage of the keyword overlap $\frac{set(a) \cap set(b)}{set(a) \cup set(b)}$, where a and b are the keywords of two leaders. We conclude that the expertise diversity (1 - keyword overlap) of heterogeneous leaders is larger than that of homogeneous leaders (Fig 4b). It suggests that heterogeneous leaders have more diverse expertise and they might be more complementary in knowledge. Our results are similar to the findings in the open source software teams, low overlap in expertise can help teams have active activities and updates (Giuri et al., 2010). Paper publishing teams and software building teams are similar since their tasks both require novelty. Innovative teams benefit more from the expertise diversity of heterogeneous leaders than the same expertise of homogeneous leaders.

*Shared Leadership and Knowledge Recency*

Usually, a paper has a list of references, which reflect the authors' depth and range of knowledge. We cannot directly differentiate the reference is cited by whom when the publication is finished by a team. Thus, we choose sole-author publications to observe researchers' citing practice. We can find that in Fig 4c-e, senior researchers have a larger reference range, reference standard deviation, and reference mean values than junior researchers. The range of references, for example, a paper published in 2020 has two references, and the earliest reference is 2010, the newest reference is 2019. Then we calculate the time distance of the reference year to the publication year, 10 and 1 separately. Thus, the range of references is 9. Similarly, we calculate the standard deviation and mean value of references. In Fig 4f-h, both senior leaders (0,0,2) have the largest reference range, standard deviation, and mean value. It verifies senior leaders can help increase the depth of papers and trace older references. Both junior leaders (2,0,0) have the smallest reference range, standard deviation, and mean values. It proves that junior leaders always keep a trace of the newest research and thus cite more recent papers. But the results tell us neither too old references nor too recent references will increase papers' impact. In Fig 4i-k, there is an inverted U-shape between reference aging and paper impact. Papers with a medium reference range have a higher citation impact than papers with both small and large ranges. This is the same for reference standard deviation and mean values. This result indicates that heterogeneous leaders with junior and senior leaders follow the newest and trace the oldest ideas, thus receiving more citations. We find the same patterns that senior researchers cite older references than junior ones (Gingras et al., 2008; Merton, 1973). Prior research suggests that when senior researchers work with junior researchers, reference aging decreases, and disruptive novelty increases (Cui et al., 2022). Our results demonstrate the assembly of junior leaders and senior leaders in shared leadership can also bring high citation influence for teams.

## Conclusion

Leadership is evolving from "individualistic, hierarchical, one-directional and de-contextualized notions" to shared leadership (DeRue, 2011, p. 125). With increasing complexity, uncertainty, and knowledge intensity in scientific tasks, a single leader cannot play all leading functions, which requires multiple leaders to participate in leadership activities. Shared leadership is a widely discussed topic in diverse creative tasks, such as R&D teams in companies (Gumusluoglu & Ilsev, 2009), open source communities (Zhu et al., 2011), and brainstorming and negotiating (Woolley et al., 2010). In this paper, we extended the concept of shared leadership to the scientific context. Scientific teams are driven by creativity and knowledge, which might be different from companies driven by profits and routine affairs. Knowledge workers have the professional pursuit and desire for leadership impact (Pearce et al., 2004). Social identity theory proposes that acting as leaders makes them feel important, have a stronger belonging sense to the group, and meanwhile boosts team innovation (Hogg, 2008). Theoretically, social identity theory helps us understand the importance of shared leadership in scientific teams. Meanwhile, we also enrich this theory by considering it in the creative scientific context. In the science of science background, we consider shared leadership as informal, rather than a strict definition of formal supervisors. We distinguish different leaders from junior, mid, to senior leaders in important positions, and six different combinations of these leaders, comparing the differences between homogeneous and heterogeneous leaders. Our main finding is that heterogeneous leaders outperform homogeneous leaders with higher citation ranks. The difference between the two leaders' career age is positively associated with the paper's high citations, even when we control time fixed effect and team size. The result that heterogeneous leaders are better is still consistent for all students of different career ages. We further explore possible mechanisms underlying this pattern. Firstly, heterogeneous leaders with both senior and junior leaders are more adaptive in large and small teams. Heterogeneous shared leadership can equip teams with both the benefits of a senior researcher (e.g., vision, impact) and a junior researcher (e.g., details, novelty). Secondly, senior leaders cite older references, whereas junior leaders follow more recent work. The complementary of the literature knowledge can build projects with depth and recency. Thirdly, senior and junior leaders have diverse expertise, whereas peer leaders have a large overlap of expertise. Heterogeneous leaders can bring heterogeneous skills and expertise to teams, which is beneficial for team performance.

Our project suggests that a combination of senior and junior leaders can maximize team performance compared to leaders of similar age. Heterogeneous teams will produce long-term, positive effects on junior scientists who want to gain leadership in science and students who are beginning their research career. How junior scientists win a seat at the decision-making table is a heated topic in science (Powell, 2021). Although junior scientists are still at the early stage of their careers, their participation in leadership can bring fresh ideas. Working with senior leaders in the form of shared leadership is a way to help junior scientists feel ownership and make up for their lack of experience. Heterogeneous leadership can boost student-centered mentorship. Exposing oneself to the perspectives of various leaders can help develop critical thinking and fuse diverging ideas (Phillips & Pugh, 2015). We still have some limitations in this paper. Paper citation is not a perfect metric for evaluating team performance (Daud et al., 2022). Authors' career ages that we calculate are not real academic ages, which would be limited by the OpenAlex dataset and author name disambiguation method. The name disambiguation might have difficulty in identifying Asian names, and thus some wrong identifications exist. We choose leaders whose career ages are above 10 and who are in important positions. The definition of leadership is relatively simple. It is possible that leaders identified in our research are still junior researchers although they have a career age of above 10, or that they are excellent talents who have risen to leadership positions at a young age. We can improve our leadership measures through cognitive, geographical, and institutional information (He et al., 2022). Also, the distinction between junior, mid, and senior based on career age is relatively arbitrary. Career age has limits in defining the experience and power of leaders. These identified leaders are not necessarily formal mentors. Instead, they are informal leaders. We will further extend our analysis to different fields and cross different disciplines as our future work to dive deeper into this landscape to understand the patterns of shared leadership and interdisciplinarity. The best way of distinguishing leaders might be the combination of career age and discipline. Furthermore, disciplinary boundaries are getting blurred, especially with the wide adoption of data science and AI. We might consider using keywords to represent the lower boundaries of disciplines. We also plan to consider the order of leaders in authorship, to see how the ranking of a senior leader or mid leader or senior leader influences team impact.

Liu, S., Hu, J., Li, Y., Wang, Z., & Lin, X. (2014). Examining the cross-level relationship between shared leadership and learning in teams: Evidence from China. The leadership quarterly, 25(2), 282-295.

Merton, R. K. (1973). The sociology of science: Theoretical and empirical investigations. University of Chicago press.

Meyer, E., & Schroeder R. (2015). Knowledge machines: Digital transformations of the sciences and humanities. MIT Press, Cambridge, Massachusetts.

Milojević, S., Radicchi, F., & Walsh, J. P. (2018). Changing demographics of scientific careers: The rise of the temporary workforce. Proceedings of the National Academy of Sciences, 115(50), 12616-12623.

Mumford, T. V., Campion, M. A., & Morgeson, F. P. (2007). The leadership skills strataplex: Leadership skill requirements across organizational levels. The Leadership Quarterly, 18(2), 154-166.

O'Toole, J., Galbraith, J., & Lawler III, E. E. (2002). When two (or more) heads are better than one: The promise and pitfalls of shared leadership. California management review, 44(4), 65-83.

Paulus, P. B., Kohn, N. W., Arditti, L. E., & Korde, R. M. (2013). Understanding the group size effect in electronic brainstorming. Small Group Research, 44(3), 332-352.

Pearce, C. L. (2004). The future of leadership: Combining vertical and shared leadership to transform knowledge work. Academy of Management Perspectives, 18(1), 47-57.

Pearce, C. L., & Conger, J. A. (2002). Shared leadership: Reframing the hows and whys of leadership. Sage Publications.

Phillips, E., & Pugh, D. (2015). EBOOK: How to Get a PhD: A Handbook for Students and their Supervisors. McGraw-Hill Education (UK).

Powell, K. (2021). How junior scientists can land a seat at the leadership table. Nature, 592(7854), 475-477.

Priem, J., Piwowar, H., & Orr, R. (2022). OpenAlex: A fully-open index of scholarly works, authors, venues, institutions, and concepts. ArXiv. https://arxiv.org/abs/2205.01833

Sperber, S., & Linder, C. (2018). The impact of top management teams on firm innovativeness: a configurational analysis of demographic characteristics, leadership style and team power distribution. Review of Managerial Science, 12(1), 285-316.
24